\documentstyle[preprint,12pt,aps,prl]{revtex}
\tightenlines
\begin{document}
\newcommand{\nvo}{NaV$_{2}$O$_{5}$}
\newcommand{\nxvo}{Na$_x$V$_{2}$O$_{5}$}
\newcommand{\lvo}{LiV$_{2}$O$_{5}$}
\newcommand{\cm}{$cm^{-1}$}
\title {Charge ordering and optical transitions of LiV$_2$O$_5$ 
and NaV$_2$O$_5$}
\author{M. J. Konstantinovi\'c$^{\ast}$, J. Dong, M. E. Ziaei,
B. P. Clayman, J. C. Irwin}
\address{Simon Fraser University, Physics Department, 8888 University 
drive, Burnaby, B.C. V5A1S6 Canada}
\author{K.Yakushi}
\address{Institute for Molecular Science, Myodaiji, Okazaki 444, Japan}
\author{M. Isobe and Y. Ueda}
\address{Institute for Solid State Physics, The University of Tokio, 
7-22-1 Roppongi, Minato-ku, Tokio 106, Japan}

\maketitle

\begin{abstract} 
We present measurements of the polarized optical spectra of {\nvo} and 
{\lvo}. In an energy range from 0.5 to 5.5 eV, we observe similar peaks in the 
$E\parallel a$ spectra of {\nvo} and {\lvo}, which suggests similar electronic 
structures along the a axis in both materials. On the other hand, we find an almost 
complete suppression of the peaks in $\sigma_b$ of {\lvo} around 1 and 5 eV. We 
attribute this suppression to the charge localization effect originating from 
the existence of a double-chain charge-ordering pattern in {\lvo}.

\end{abstract}


In the past several years, quantum phenomena resulting from the low 
dimensionality of effective electron interactions in solids have been 
investigated with increasing intensity from both experimental and theoretical 
points of view.
The increase in interest was partially motivated by the discovery of 
inorganic materials which exhibit quantum effects, such as the $Sr-Cu-O$ system 
\cite {a1} or $CuGeO_3$ \cite {a2}, and by a common belief that these studies 
would give us a better understanding of electron correlations in general.

The vanadate family of $AV_2O_5$ oxides have demonstrated a variety of the 
low-dimensional phenomena which originate from their peculiar crystal structures 
\cite {a4}. These oxides are quasi two-dimensional (2D) materials with layers 
formed by $VO_5$ square pyramids. The A atoms are situated between layers as 
intercalants, but in fact they determine the valence state of vanadium atoms 
(acting as charge reservoirs). If the A atoms belong to the first column in the 
periodic table, such as A = Li, Na, each valence electron is shared between two 
vanadium atoms. As a result the V ions are in a mixed valence-state with an 
average valence of +4.5. The common consequence of mixed valence in these 
structures is the appearance of a quasi-1D magnetic interaction, since chains 
carrying the spin (made of V$^{4+}$, S=1/2) are separated from each other by 
nonmagnetic chains (V$^{5+}$). In both {\lvo} and {\nvo} the 1D character of the 
magnetic ordering was confirmed \cite {a5,a6}. In addition, there is a 
possibility of the existence of strong valence fluctuations, and eventually 
charge ordering (CO) effects.  

A very interesting interplay between spin and charge dynamics results in the 
phase transition discovered in NaV$_2$O$_5$ \cite {a5}. Up to now, accumulated 
experimental data \cite {a5,a7,a8,a9,a10} suggested that {\nvo} exhibits the CO 
phase transition at T=34 K into a gapped spin-liquid ground state. The arguments 
are mostly based on the insensitivity of the phase-transition effects associated with 
magnetic fields. Subsequently, several theoretical analyses of the role of the 
electron correlations (intersite Coulomb interaction) in charge dynamics and/or 
charge ordering of {\nvo} were presented \cite {a11,a12,a13,a14}. In these 
studies, the various charge-ordering ground states were proposed for the 
low-temperature phase of {\nvo}. These concepts were tested by comparison with 
optical conductivity data \cite {a15,a16,a17} with some success, but no  
consistent picture has yet emerged. 
These concepts were tested by comparison with 
optical conductivity data \cite {a15,a16,a17,a17a} with some success but no  
consistent picture has emerged yet. 
None of the models proposed to date have reproduced optical transitions in the
0.8-5.5 eV range and provided an explanation for the origin of the low-frequency
excitations (observed in both IR and Raman spectra)\cite {a15,a10}.
In fact, the central issue refers to the energy scale at which the CO in {\nvo} 
should manifest itself, and what should be a fingerprint of it in the optical 
spectra.

In {\lvo} the effects caused by the uniform vanadium valence are not observed 
\cite {a18}, and the structure is assumed to be in a charge-ordered phase 
(without a spin gap) even at room temperature \cite {a19}. We have measured and 
compared the optical spectra of {\lvo} and  {\nvo}. On the basis of these 
results, we discuss the origin of optical excitations and CO ground states in 
both compounds, and the nature of the CO phase transition in {\nvo}.

Single crystals, with dimensions typically 1x3x0.2 mm ({\nxvo}) and 2x3x1 
({\lvo}) along a, b, and c axes respectively, prepared as described in Ref. 
\cite {a20} were studied. The measurements were performed on (001) surfaces.
Measurements of the polarized infrared reflectivity spectra were performed on 
Fourier transform spectrometer Bruker IFS 113V, in an energy range from 40 to 
11000 \cm. An {\it in situ} overcoating technique was used in reflectivity 
measurement \cite {a20a}. 
The reflectivity in the range from 1100 to 3300 \cm was measured on
an Atago Multiviewer spectrometer with multichanel detection
system combiend with a SPECTRA TECH IR-Plan microscope.
A rotating-analyzer ellipsometer was used to measure the pseudodielectric 
function in 1.2-5.5 eV energy range.

At room temperatures the $\alpha' -${\nvo} and $\gamma -${\lvo} have 
orthorhombic unit cells \cite {a7,a21} (described with space groups $Pmmn$, and 
$Pnma$, respectively), and crystal structures consisting of layers of VO$_5$ 
square pyramids which are mutually connected via common edges and corners making 
the characteristic V "zigzag" chains along the b axis. 
Parameters a and b, are similar in both compounds \cite {a7,a21}. The c axis 
of {\lvo} is approximately twice as large, since the {\lvo} unit cell comprises four 
formula units (two in {\nvo}). In {\nvo} all vanadium atoms are in uniform 
valence state at room temperatures (an average valence of +4.5), and thus 
indistinguishable in the unit cell (i.e. they occupy sites with same symmetry).

Conversely, the structure of {\lvo} is characterized by two kinds of 
vanadium chains along the b axis. One is magnetic, V$^{4+}$ (S=1/2) and the 
other nonmagnetic V$^{5+}$ (S=0), see Fig.1. Another important difference in 
crystal structures comes from the different sizes of Li and Na ions. Li atoms 
are smaller, and consequently the VO layers in {\lvo} are more corrugated, see Fig.1.
An alternative description of the $VO_5$ layer is that it consists of V-O-V 
rungs coupled together in a ladder fashion through the oxygen bonds along 
the b axis. These ladders are mutually connected to each other via a direct 
overlap of vanadium d orbitals along the $V^1_R-V^2_L-V^1_R$ "zigzag" chain, 
see Fig.1.

The optical conductivity of {\nvo} and {\lvo} is presented in Fig.2. The optical 
conductivity is calculated from the reflectivity data using Kramer-Kronig 
relations. The pseudodieletric functions of {\nvo} and {\lvo} are shown in Fig.3
The 1.5-5.5 eV energy range is computed using the ellipsometric equations for 
the isotropic case. Consequently, $\epsilon$ ($\sigma$) represents a complicated 
average of the projections of the dielectric tensor on the sample surface. We 
present the spectra of the (001) surface taken with the a axis (thin line) and 
the b axis (thick line) in the plane of incidence. Following Aspnes.s 
prescription, \cite{a22} we attribute these components to the dielectric tensor
components $\epsilon^{2}_{aa}$ ($\sigma_a$) and $\epsilon^{2}_{bb}$ 
($\sigma_b$), respectively.

Bands with energies at 0.9, 1.2, 3.22, 4.23, and 5 eV for $\sigma_a$ and 1.1, 1.58, 
3.73, and 5 eV for $\sigma_b$ are found in {\nvo}; see Fig. 2a. The same structures were 
observed in previous studies \cite {a15,a16,a17,a17a} as well.
In {\lvo} we find bands with energies centered at 0.85, 3.03, 4.20, and 4.95 eV 
for $\sigma_a$ and at 3.42 eV for $\sigma_b$. It is important to note that 
while $\sigma_a$ in {\lvo} closely resembles $\sigma_a$ in {\nvo}, 
$\sigma_b$ in {\lvo} is almost completely suppressed except 3.42 eV mode.

We first focus on excitations around 1 eV in the {\nvo} spectra, and discuss the 
results in light of the electronic band structure of {\nvo} obtained from 
density-functional calculations, (DFC's) \cite {a23} and t-j-V model 
\cite{a11,a12,a13,a14}. According to DFC's, the vanadium d-level degeneracy is 
removed due to anisotropy of the crystal field \cite {a23} and the lowest 
occupied $3d_{xy}$ states are separated by 1-5 eV from remaining 3d states.
This energy scale provoked the assignment of 0.9 eV peak in the optical spectra, 
Fig.2, as a transition between d-d crystal-field levels of vanadium ions \cite 
{a16}. However, recent work on Ca-doped {\nvo} showed that 0.9 eV peak 
decreases in intensity with increasing Ca, \cite {a24} 
This result seams to be inconsistent with the d-d transition picture (the d-d
transition intensity should be proportional to the number of 
$V^{4+}$ ions). On the other hand, in the t-J-V model, the combined effects of 
the short-range Coulomb interaction and valence fluctuations of V ions determine 
the peak energies in the optical conductivity spectra, e.g. the anisotropy of the 
interband transitions in the a and b directions. In order to reproduce 
experimental observations, basically all t-J-V calculations rely on (or predict) 
the existence of strong charge discomensuration, which is not in accordance with 
effects related to the uniform valence in {\nvo}.

In fact, the quarter-filled nature of the V-O-V rung \cite {a23}(0.5 electrons 
per vanadium site) suggests that the band states are superposition of the 
$d_{xy}$ molecular orbitals of bonding and antibonding types. Then it can 
argued that 0.9 eV structure corresponds to the bonding-antibonding transition 
within the V-O-V rung \cite {a15}. The energy separation of the 
bonding-antibonding $d_{xy}$ orbitals, according to the Hubbard model of the isolated rung, 
is $\Delta E_{BA} \sim 2t_a$. A reasonable value of $t_a = 0.45 $ eV reproduces 
the energy band at 0.9 eV in $E \parallel a$ spectra. Such an analysis predicts
the existence of a 
similar structure along the b-direction as indeed observed in the 1.2 eV peak in  
$\sigma_b$.

The temperature dependence of the optical conductivity raises even more 
questions. All the features in $\sigma_a$ spectra increase in intensity,
but without a change of energy at CO phase transition temperature, $T_c=34 K$, 
\cite {a17,a24a}. According to the t-j-V model, strong energy dependence of 0.9 
eV structure is expected across the phase-transition temperature, since the 
driving mechanism for the CO is short-range Coulomb interaction 
(which induces a non zero in-rung charge disproportion potential
 \cite {a12,a15}). Switched on at
$T_c$, this interaction naturally produces "zigzag" charge order. Thus one may 
either conclude that change of charge disproportion below phase transition is 
very small \cite {a24a} [this contrasts with the strong splitting of V NMR lines 
observed below $T_c$ (Ref.7)] or that CO does not manifests itself
through the change of energy of 0.9 eV peak.

With this in mind, let us now discuss the optical conductivity of {\lvo}.
If we assume the bonding-antibonding transition (with and/or without the charge 
disproportion potential $\Delta$) to be responsible for the 0.9 eV 
optical excitation in $\sigma_a$ spectra of {\nvo}, the existence of a similar 
structure (0.85 eV) in {\lvo} at first seems to be completely unexpected. The 
reason for this is the existence of plane corrugation and strong double chain charge 
ordering in {\lvo}. However, despite corrugation, the $3d_{xy}-2p-3d_{xy}$ bonds 
of the rungs in these two compounds are similar. According to the crystal  
structures, the V-O-V bond angles differ in these two structures by $10^o-20^o$
($120^o$ in {\lvo} and $140^o$ in {\nvo}); see Fig.1. Such a structural 
difference would eventually cause a somewhat smaller $t_a$ hopping in {\lvo}. If we 
discard $\Delta$, we find $t_a \sim 0.42 eV$ in {\lvo}. On the other hand, the 
double-chain charge order in {\lvo} should give completely different $\Delta$ then in 
{\nvo}, which should cause large difference in the optical conductivity, which 
is not observed. So, whether or not the difference in charge disproportion 
between these compounds manifests itself in $E\parallel a$ peak energies is still an 
open question. 

Intensity estimates are much more difficult to perform. In the simplest 
approach, the changes in intensity are produced by different hoppings. 
In {\nvo} 
the $E\parallel a$ peak is around three times more intense than the 
$E\parallel b$ peak. 
Thus, according to the t-J-V model \cite{a14}, the hopping energy $t_b$ is expected 
to be at least two times smaller, $t_b \sim 0.2$ eV. 
From {\lvo} optical spectra we learned that major effect of charge localization 
involves peak intensities. As we already discussed, {\lvo} is at room 
temperature in a charge-ordered state, i.e. a double-chain charge ordering of 
electrons along the b axis, see Fig.1. In this case the electronic transitions 
along b axis to a states with double-site occupancies should be almost 
completely suppressed. This is evident by vanishing of the structures around 1 
eV in $E \parallel b$ spectra of {\lvo}. This effect is caused by a reduced  
probability for the electrons to hope along the b axis 
or in the xy direction that are already occupied. This is consistent with 
the vanishing of the structures around 1 eV in the $\sigma_b$ spectra of {\lvo}.

The $E \parallel a$ bonding-antibonding transition is not influenced by the
double-chain charge ordering pattern in {\lvo} as much as processes described 
above, and we still find the peak at 0.85 eV. Its intensity is approximately two 
times smaller than the 0.9 eV peak in {\nvo}, indicating that 
charge-localization also affect (in some way) this process. If so, suppression of the 
0.9 eV peak in {\nvo} is expected below the phase transition and indeed 
observed in Ref. 27.

Therefore, the intensities of the peaks along the b axis in the optical conductivity
of {\nvo} should be strongly temperature dependent if the CO below $T_c$ is of 
"in-line" type. This is not observed in the experiment, firmly establishing the 
"zigzag" CO scenario in {\nvo}, \cite {a11}.

Keeping in mind the complete disappearance of the 1.1 and 1.6 eV peaks in the 
$\sigma_b$ spectra of {\lvo}, we propose that these two structures in {\nvo} 
originate from electronic transitions which involve double-electron occupation 
of the rungs created in neighboring ladders or the same ladder, respectively; see 
Fig.1. That is, the $E \parallel b$ experimental configuration allows both intraladder 
and interladder transitions, while the $E \parallel a$ configuration allows 
only interladder 
transitions. Thus the interladder transitions could correspond to 1.1 eV 
peak (Fig. 2), 
which has a similar intensity in both $\sigma_a$ and $\sigma_b$ spectra. 
Different energies of intraladder and interladder transitions could be related to the 
Coulomb potential difference in the following way: Let us assume that 
double-electron occupancy 
costs an effective energy V for the isolated rung. Then the 
total potential difference between these two cases is (taking $V_{xy} \sim \sqrt 
2 V$) $\Delta E =(2V+2V_{xy})-(3V+V_{xy})= \sim 0.4 V$. Taking V=1 eV, we obtain 
$\Delta E \sim 0.4 $eV. Since experiment gives an energy difference
of about 0.5 eV, the additional energy difference between of about 0.1 eV 
could be due to the difference in hopping. 
If so, the first consequence is that interladder hopping $t_{xy}$ is 
not an order of magnitude smaller than $t_b=0.23 $ eV ($J_b=4t_b^2/U$ 
$J \sim 560 K$), \cite {a6}, but rather just factor of two or three smaller,
 $t_{xy} \sim 0.1 eV$. Such a conclusion is consistent with 
previous estimates \cite {a25} and with arguments involving
magnetic dimer formation along the xy direction, which follows from the 
"zigzag" charge-ordered ground state \cite {a11}.

The structures around 3-4 eV do not show much difference in these two
compounds. According to the angle-resolved photoemission
spectroscopy  results \cite {a22a},
we assign 3.22 and 3.73 eV peaks to $O_{2p} - V_{3d}$ transitions
within the same V-O-V rungs. 

In conclusion, we studied the electronic properties of {\nvo} and {\lvo}
by measuring the optical reflectivity and dielectric functions
of these two compounds in the 0.5 - 5.5 eV energy range. While 
$\sigma_a$ is similar in both compounds the $\sigma_b$ is strongly suppressed
around 1 and 5 eV in {\lvo}. We atribute this effect to charge-localization
originating from the double in-line charge-ordering pattern in {\lvo}.
Our results, thus, support the zigzag charge-ordering ground state in
{\nvo} below the phase-transition temperature.

Acknowledgement:
This work was supported by the Natural Sciences and Engineering Research
Council of Canada, and Simon Fraser University.
$^{\ast}$ mkonstan@sfu.ca

\bibliographystyle{prsty}

\begin{thebibliography}{99}
\bibitem{a1} E. Dagotto, T. M. Rice, Science {\bf 271}, 618 (1996).
\bibitem{a2} M. Hase, I. Terasaki and K. Uchinokura, Phys. Rev. Lett.
{\bf 70}, 3651 (1993).
\bibitem{a4} Y. Ueda, M. Isobe, J. Magn. Magn. Materials {\bf 177-181},
741 (1998).
\bibitem{a5} M. Isobe and Y. Ueda, J. Phys. Soc. Jpn. {\bf 65},
1178 (1996).
\bibitem{a6} M. Isobe and Y. Ueda, J. Phys. Soc. Jpn. {\bf 65},
3142 (1996).
\bibitem{a7} H. G. von Schnering, Y.Grin, M. Kaup, M. Somer, R. Kremer,  
O. Jepsen, T. Chatterji and M. Weiden, Zeit. Kristall. {\bf 213}, 246 (1998). 
\bibitem{a8} T. Ohama, H. Yasuoka, M. Isobe and Y. Ueda, Phys. Rev B {\bf 59},
3299 (1999).
\bibitem{a9} M. Weiden, R. Hauptman,
C. Geibel, F. Steglich, M. Fischer, P. Lemmens and G. G{\"u}nterodt,
Z. Phys. B {\bf 103}, 1 (1997).
\bibitem{a10} Z.  V.  Popovi\'c  M. J. Konstantinovi\'c,
R. Gaji\'c, V. N. Popov, Y. S. Raptis, A. N. Vasil'ev, M. Isobe and  
Y., Ueda, J.  Phys.  Cond.  Matter {\bf 10}, L513 (1998).
\bibitem{a11} H.  Seo and H. Fukujama, J. Phys. Soc. Japan {\bf 67}, 2602
(1998).
\bibitem{a12} S. Nishimoto and Y. Ohta, J. Phys. Soc. Jpn {\bf 67}, 36
(1998).
\bibitem{a13} P. Thalmeier and P. Fulde, Europhys. Lett. {\bf 44}, 242 (1998).
\bibitem{a14} M. Cuoco, P Horsch and F Mack, Phys. Rev. B {\bf 60}, R8438
(1999).
\bibitem{a15} A. Damascelli, D. van der Marel, M. Gruninger, C. Presura,
T. T. M. Palstra, J. Jegoudez, and A. Revcolevschi, Phys Rev Lett.
{\bf 81}, 918 (1998).
\bibitem{a16} S. A. Golubchik, M. Isobe, A. N. Ivlev, B. N. Mavrin, M. N.
Popova,
A. B. Sushkov, Y. Ueda and V. N. Vasil'ev, J. Phys. Soc. Jpn {\bf 66}, 4042
(1997).
\bibitem{a17} M. J. Konstantinovi\'c, L. F. Lastras-Martinez, M. Cardona,
Z. V. Popovi\'c, A. N. Vasil'ev, M Isobe and Y. Ueda, Phys. Stat. Sol. (b)
{\bf 211}, R3 (1999).
\bibitem{a17a} V. C. Long, Z. Zhu, J. L. Musfeldt, X. Wei, H. -J. Koo,
M. -H Whangbo, J. Jegoudez, and A. Revcolevschi, Phys. Rev. B {\bf 60}, 15721
(1999).
\bibitem{a18} Z. V. Popovi\'c, R. Gaji\'c, M. J. Konstantinovi\'c,
R. Provoost, V. V. Moshchalkov, A. N. Vasil'ev, M. Isobe and Y. Ueda,
Phys. Rev. B {\bf 61}, 11454 (2000).
\bibitem{a19} Y. Takeo, T. Yosihama, M. Nisji, K. Nakajima, K. Kakurai,
M. Isobe
and Y. Ueda, J. Phys. Chem. Solids {\bf 60}, 1145 (1999).
\bibitem{a20} M. Isobe and Y.  Ueda, J. of Alloys and Comp. {\bf 262}, 180
(1997).
\bibitem{a20a} C. C. Homes. M. Reedyk, D. A. Crandles, and T. Timusk, Appl. Opt.
{\bf32}, 2976 (1993).
\bibitem{a21} D. N. Anderson and R. D. Willett, Acta Crystallorg. Sec B:
Struct. Crystallogr. Cryst. Chem. {\bf 27}, 1476 (1971).
\bibitem{a22} D. E. Aspnes, J. Opt. Soc. Am. {\bf 70}, 1275 (1980).
\bibitem{a22a} K. Kobayashi, T. Mizokawa, A. Fujimori, M. Isobe and Y. Ueda,
Phys. Rev. Lett. {\bf80} 3121 (1998).
\bibitem{a23} H. Smolinski, C. Gros, W. Weber, U. Peuchert, G. Roth,
M. Weiden
and C. Geibel, Phys. Rev. Lett. {\bf 80}, 5164 (1998).
\bibitem{a24} C. Presura, D. van der Marel, M. Dischner, C. Geibel
and R. K. Kremer, cond-mat/0005536 (unpublished).
\bibitem{a24a} C. Presura, D. van der Marel, A. Damascelli and
R. K. Kremer, Phys. Rev. B {\bf 61}, 15762 (2000).
\bibitem{a25} V. A. Ivanov, Z. V. Popovi\'c, O. P. Khuong,
and V. V. Moshchalkov, cond-mat/9909046 (unpublished).

\end{thebibliography}

\begin{figure}
\caption
{Schematic representation of the (a) LiV$_2$O$_5$
and (b) NaV$_2$O$_5$ crystal structures.}
\label{fig1}
\end{figure}
\begin{figure}
\caption
{Room temperature optical conductivity of {\nvo} and {\lvo}.}
\label{fig2}
\end{figure}
\begin{figure}
\caption
{Imaginary part of the pseudodielectric function of {\nvo} and {\lvo}.}
\label{fig3}
\end{figure}

\end{document}